# Simulation of Conventional Droop Controller for Islanding DGs

Khodakhast Nasirian[1], Hadi Taheri[1]

**Khodakhast Nasirian \***, Ph.D. student in Electrical Engineering Department, Marvdasht Azad University, Marvdasht, Iran

**Hadi Taheri** Masters student. student in Electrical Engineering Department, Marvdasht Azad University, Marvdasht, Iran

## I. ABSTRACT

A microgrid is a new concept that has changed the power systems dramatically. It is a combination of Distributed Generation Resources (DGR) like Biomass, PV systems, Wind energy, Fuel cell, Diesel Generator, and so on with different types of loads (residential or commercial). Microgrids can work in two modes: autonomous and Interconnected. In the Islanding situation, the loads will be supported by DGR and without connecting to upstream utility grids. Controlling power electronic Interfaces between sources and loads has been an important task between the researchers. Several different strategies have been presented by researchers. Droop control strategy is one of the which has its pros and cons. In this paper, the conventional droop strategy has been explained in detail and formulated. The Simulation results are taken from MATLAB/SIMULINK to show the capability of the control strategy.

*Keywords: Microgrid, Distributed Generation Resources (DGR), Islanding situation, Control strategies, Droop control*

## II. INTRODUCTION

By increasing the dependence of modern life on Electric equipment and computer systems, power quality and reliability are two essential needs. This need for critical loads such as hospitals, -Telecommunication systems, and information centers is felt more and more. On the other hand, the production of power at the centralized power plants and its transmission face many problems, such as environmental pollution, occupation a lot of ground for transmission lines, and voltage drop Which causes the huge cost of electricity to be consumed. According to these facts, in recent years the necessity for considering other technologies for generating electricity which need less investment with better quality and reliability has been quite tangible. Recent Developments in Small-scale power generation technologies and utilization Renewable Energies such as photo-voltaic as well as innovation-In power electronics, it causes a high tendency among power companies to explore of Distributed Generation Recourses (DGR) in the distribution system and Near to consumers [1-9].

For continuous feeding of sensitive loads and use other benefits of DGR, a concept that has been called the Microgrid has emerged. Figure (1) shows a defined microgrid. Microgrid components Include DGR, sensitive loads, and energy storage devices, and it works in two modes: interconnecting to the grid and independent mode (islanding). A lot of research such as planning, optimization, power quality improvement, have been conducted with this new concept. Meanwhile, other research such as investment methods, cost, forecasting, political aspects, and so on should be taken more seriously for microgrids. some software is using for microgrids simulation such as MATLAB, PSCAD, and LTSpice [10-13]. Controlling Microgrid is the most important challenge for getting the most advantages of them. The main purpose of control in a Microgrid is preserving two major parameters voltage and frequency of the Microgrid in the allowed range and supplying the required power for loads. Other goals such as power quality, reliability, or proper sharing of power may also be other attractive goals. For better control, Distributed Generations connect to Microgrid with inverter interfaces and therefore the main purpose of control is finding the applicable strategy for controlling inverters. Several methods for controlling interface inverters have been proposed which Master/Slave method, Droop control, and difference with the average amount of power can be named here.

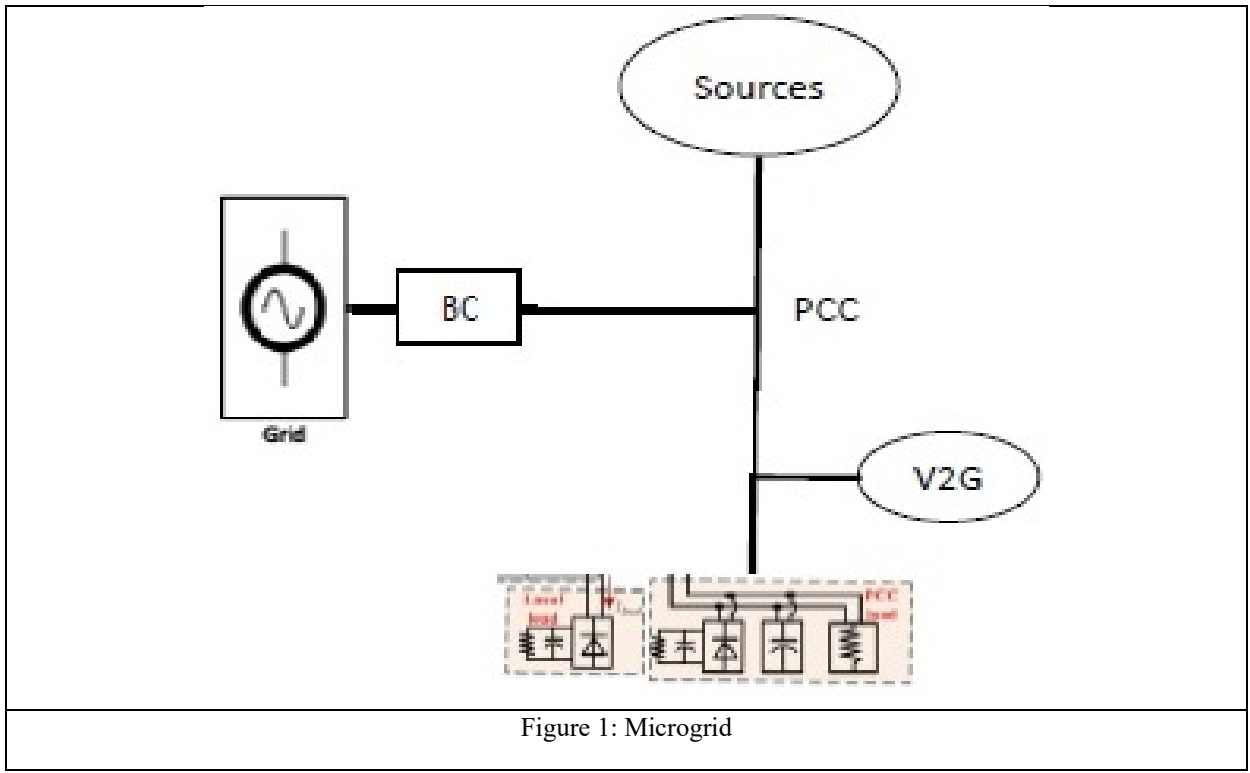

Figure 1: Microgrid

In this project, a control plan base on the droop control technique for islanding mode is proposed which will be able to meet all the control objects. The main advantage This suggested method is that this method does not need a Master unit for controlling voltage and switching between inverter controllers in isolated mode. Because only one controller for all operating modes is used. The conventional droop strategy has been explained in detail and formulated. The Simulation results are taken from MATLAB/SIMULINK to show the capability of the control strategy.

### III. MICROGRID SYSTEM

Figure (1) illustrates the arrangement of a micro-grid Includes three Distributed Generation units with inverter interfaces, a storage unit, and a sensitive load. A microgrid has been connected to the distribution network through an isolator breaker. During normal operation, the Microgrid is a part of the Distribution network. In this condition, the distribution network maintains the bus voltage and frequency of the system. Therefore, in This mode, the goal of controlling the inverter is to generate a constant value of Active and Reactive power that is named PQ control. When a fault occurs in the distribution network, the breaker in point coupling has been disconnected and the Microgrid operates independently and thus ensures Increasing reliability in the system. In this mode, the inverter controller must work in such a way that the load voltage and frequency of the microgrid in one Acceptable scale around nominal values and needed power for demand load should be provided [1-4].

In independent mode, another important issue is the proper allocation of the required power to the micro-grid loads between generation units. In other words, the purpose of the control in the DGs is to share the shift of load variation along with maintaining voltage and frequency. Based on conducted research the droop controller is one of the most effective methods for synchronization Power generation among several generators because they are capable to Set up power output for system stability quickly. Besides, the droop does not require a communication system between Distributed Generation units. A Transmission line with negligible resistance, as in figure (2), active and reactive powers injection equations can be written as follows:

$$P = \frac{V_1 V_2}{X} \sin\alpha \quad (1)$$

$$Q = \frac{V_1}{X} - \frac{V_1 V_2}{X} \cos\alpha \quad (2)$$

That is here, P is active injection power, Q reactive injection power to the line, δ power angle, $V_1$, and $V_2$ are voltages on both sides of the line and, X is the transmission line's reactance.

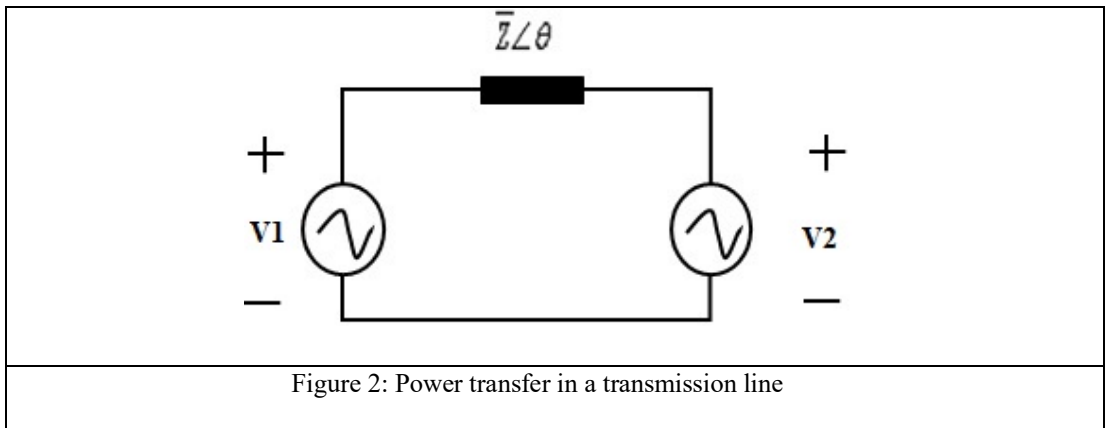

Figure 2: Power transfer in a transmission line

For the small value of δ, sin δ = δ and cos δ = 1. So, the above equalities can be reworked in terms of power and voltage angle:

$$\delta = \frac{XP}{V_1 V_2} \quad (3)$$

$$V_1 - V_2 = \frac{XQ}{V_1} \tag{4}$$

equations (3) and (4) indicate that the δ alters directly with the active power while the voltage variation differs on reactive power. In other words, controlling the active power angle will be regulated, and similarly, with reactive power regulation, the voltage also can be controlled. Since the generation units do not accept the primary period quantities of new components are independent, in droop method, each generation unit uses frequency as an alternative of power angle to regulate transitional active power. Frequency management vigorously controls the δ and similarly, the Reactive power regulation can also control the voltage. These results lead to droop control equations for the frequency and voltage.

$$f = f_0 - k_{pf}(P - p_0) \tag{5}$$

$$V_1 = V_0 - k_{vf}(Q - Q_0) \tag{6}$$

Where $K_{qv}$ the and $k_{pf}$ the droop control parameters and both are positive. $F_0$ and $V_0$ nominal frequency and voltage correspondingly, $P_0$ and $Q_0$ are default setting goals for active and active power. Equation (5) can also be used in terms of power is:

$$\delta = \delta_0 - k_{p\delta}(P - p_0) \tag{7}$$

That is here $\delta_0$, Base power angle, $K_{p\delta}$ is the droop control parameter. The results obtained from the drop method represent the voltage and frequency (or power angle) that must be provided by the inverter to be produced. The characteristics diagram of droop control is shown in Figures (3) and (4).

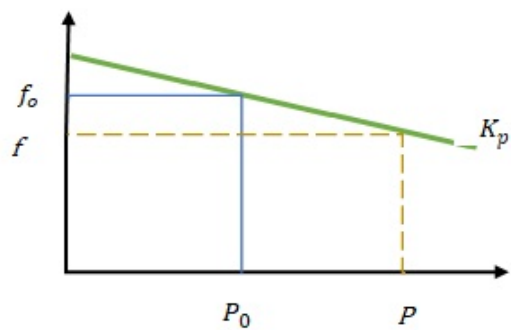

Figure (3): Characteristic Frequency-Active Power

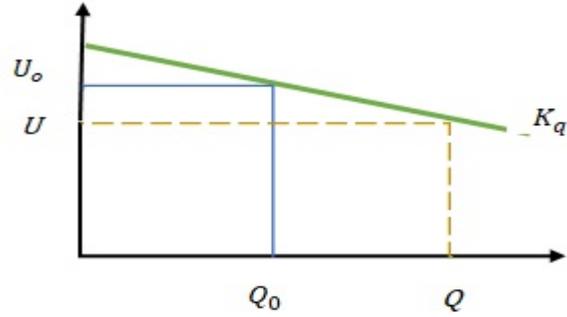

Figure (4): Voltage-reactive power droop characteristic

## IV. CONTROLLING DG'S WITH INVERTER INTERFACE IN MICROGRID

Fig. (5) Shows the proposed control plan diagram for an inverter. As mentioned earlier, the main goal of the inverter controller here is to sustain the voltage and frequency of the Microgrid and especially sensitive loads within the allowable limits. The sharing of load variation between DG's in microgrid It is also a secondary objective. These goals are considered in both modes of the microgrid, namely connected mode to the network and islanding. The proposed control method in this project has three sections which are composed of: voltage and frequency control, active and reactive power control and signal generation sector for inverter gate. The controller is active in both operation modes. Therefore, unlike other methods, no need to have a master unit or central controller for having a connection between generation units. There is also no need to switch and change the controller at the time of islanding [5-7].

### VOLTAGE AND FREQUENCY CONTROL

To monitor the load voltage and frequency of the microgrid, droop control with Frequency-active power and reactive power-voltage is used. Therefore, it is assumed that the inductance in transmission lines is, much larger than their resistance. In this part, inputs are frequency and voltage in inverter output, and measured and output powers are the default for DG. The characteristics of the F and V droop are applied to produce the active output power of each unit:

| | |
|---|---|
| $P_{ref} = P_0 - k_{fp}(f_0 - f)$ | (7) |
| $Q_{ref} = Q_0 - k_{vq}(V_0 - V)$ | (8) |

Power in each DG. V and F are frequency and voltage effective values in output bus Kpf and Kqv are droop coefficients. Coefficient and Pref and Qref are modified reference active and reactive values. In the interconnected approach, the bus voltage and frequency of the system will be saved by the nominal values of the distribution system. Therefore, generation units generate the default power values.

| $f = f_0 \Rightarrow P_{ref} = P_0$ | (9) |
|---|---|

| $v = v_0 \Rightarrow Q_{ref} = Q_0$ | (10) |
|---|---|

And voltage determined reference active and reactive powers.

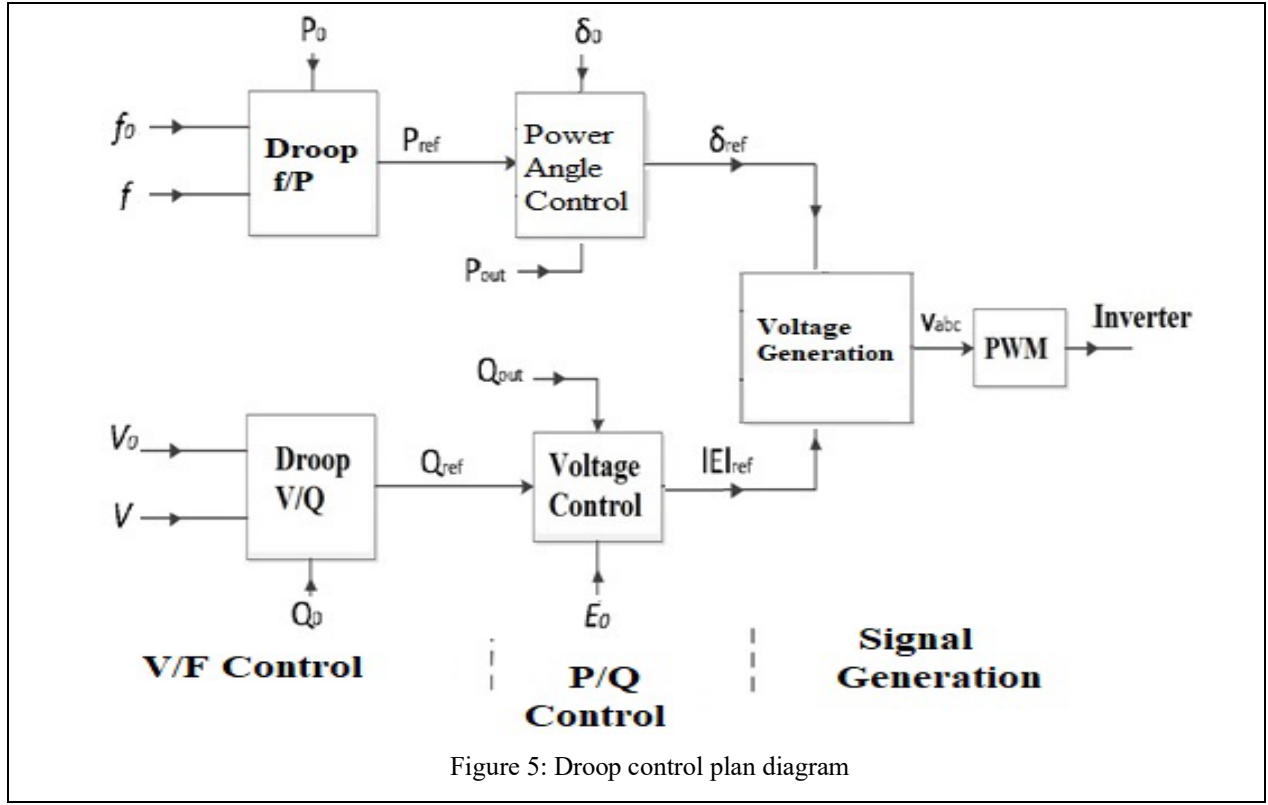

Figure 5: Droop control plan diagram

## V. ACTIVE AND REACTIVE POWER CONTROL

The purpose of this part of PQ control is to generate desire output powers. This part of control just uses local measured powers. Fig (15) shows a simple model of Micro generator. Droop equations for this generator can be written as below:

| $\delta_{ref} = \delta_0 - k_{p\delta}(P_{ref} - P_{out})$ | (11) |
|---|---|

| $E_{ref} = E_0 - k_{qE}(Q_{ref} - Q_{out})$ | (12) |
|---|---|

$P_{ref}$ and $Q_{ref}$ are reference power values which were calculated in the previous part. Pout and Qout have measured the output power value of the inverter. $K_{p\delta}$ and $k_{qE}$ are droop coefficients for voltage and power angle. $\delta_0$ and $E_0$ are default values of power angle and magnitude. These values are chosen which in interconnected mode active and reactive power from the inverter are close to default values ($Q_0$ and $P_0$). With determining E and δ can control active and reactive output.

## VI. GENERATION OF THREE-PHASE REFERENCE VOLTAGES

Generating three-phase reference voltage is taken place with $E_{ref}$ and $\delta_{ref}$ with the below equations.

$$V_a = E_{ref} \sin(\omega t + \delta_{ref})$$
$$V_b = E_{ref} \sin(\omega t + \delta_{ref} + 120^o)$$
$$V_c = E_{ref} \sin(\omega t + \delta_{ref} - 120^o)$$
(13)

And three-phase reference voltages will be transferred to gate signals by PWM [7-9].

## VII. SIMULATION RESULTS

All characteristics of the proposed system are given in Table (1). Figure (6) shows the full design of the system with two parallel-inverters for the conventional droop. As mentioned before the poor transient performance, reducing voltage and reactive power are some problems which the conventional droop cannot face. The simulation results are presented in Figure (5) and (6). In figure (5) poor performance of controller during load changing (between 1-2 (s)) is shown. It takes a long time for the controller to beat that oscillation of active and reactive powers.

Table- I: Characteristics of the designed system

| | |
|---|---|
| Vrms | 400 |
| Load | 10 KW, 5Kvar and 20KW and 10 Kvar |
| Frequency | 60 |
| R line | 0.5 |
| Active Power DG 1 | 5 KW, and 10KW |
| Active Power DG 2 | 5 KW, and 10KW |
| Reactive Power DG 1 | 2.5Kvar a and 5 Kvar |
| Reactive Power DG 2 | 2.5Kvar a and 5 Kvar |

The simulation results are presented in Figure (7-10). In figure (7) poor performance of controller during load changing (between 1-2 (s)) is shown. It takes a long time for the controller to beat that oscillation of active and reactive powers. As shown in figure (8-10) during a sudden change in load, the voltage drops and causes that the active and reactive power on the load could not be fed completely. For a better understanding of the drop voltage during the load changing, the RMS voltages from inverters and load are shown in figure (7).

## VIII. Conclusion

In this research, a conventional control for droop strategy for applying to DGR in parallel connection was presented. The initial equation for achieving the final equations was given based on two parallel synchronous generators. It was shown and proven that this kind of droop can be applied to new grids. The advantages and disadvantages of the proposed controller were explained. In the end, The Simulation results are taken from MATLAB/SIMULINK to show the capability of the control strategy.

| 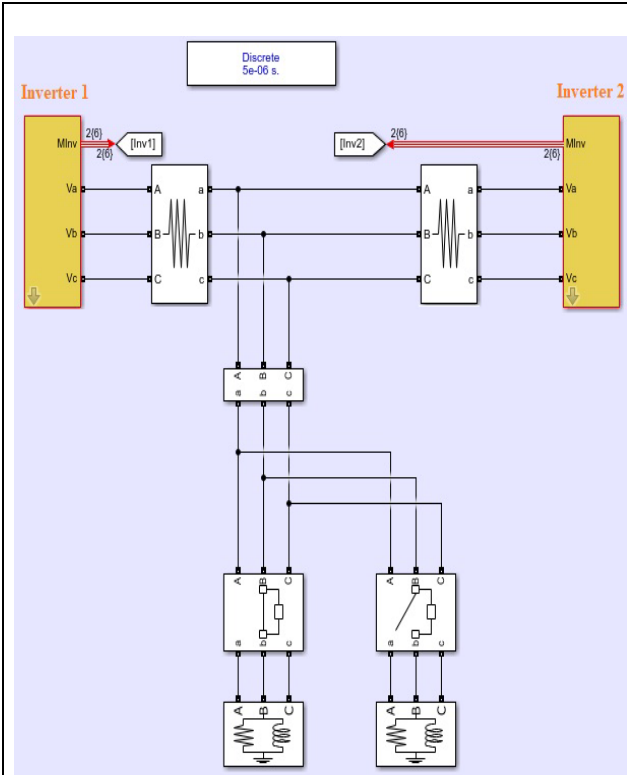 | 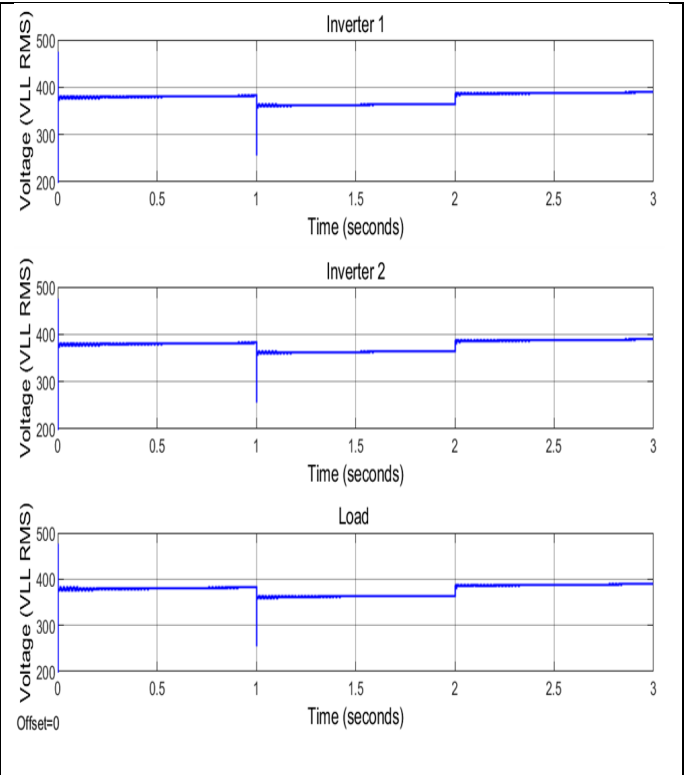 |
|---|---|
| Figure 6: Droop Control Simulation | Figure 7: Vrms |

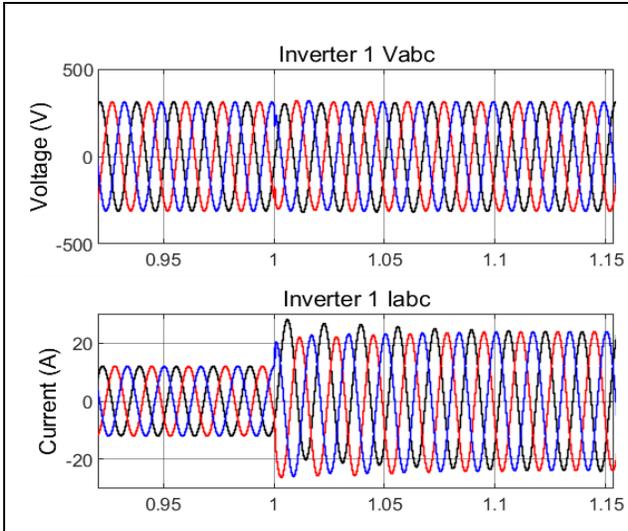

Figure 8: Voltage and current in DG1

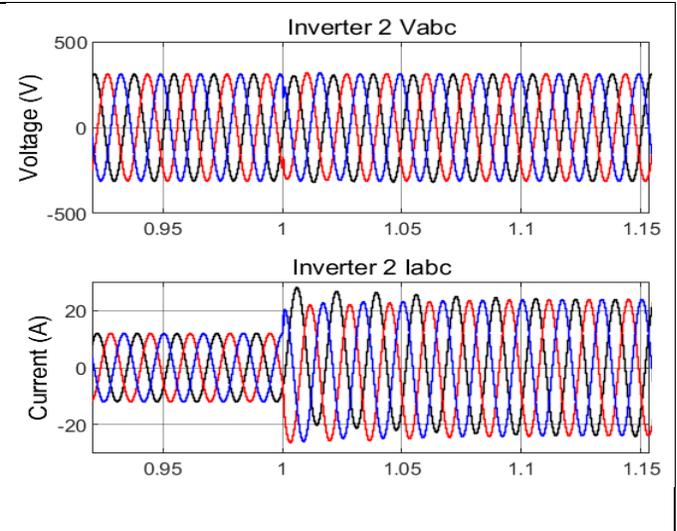

Figure 9: Voltage and current in DG2

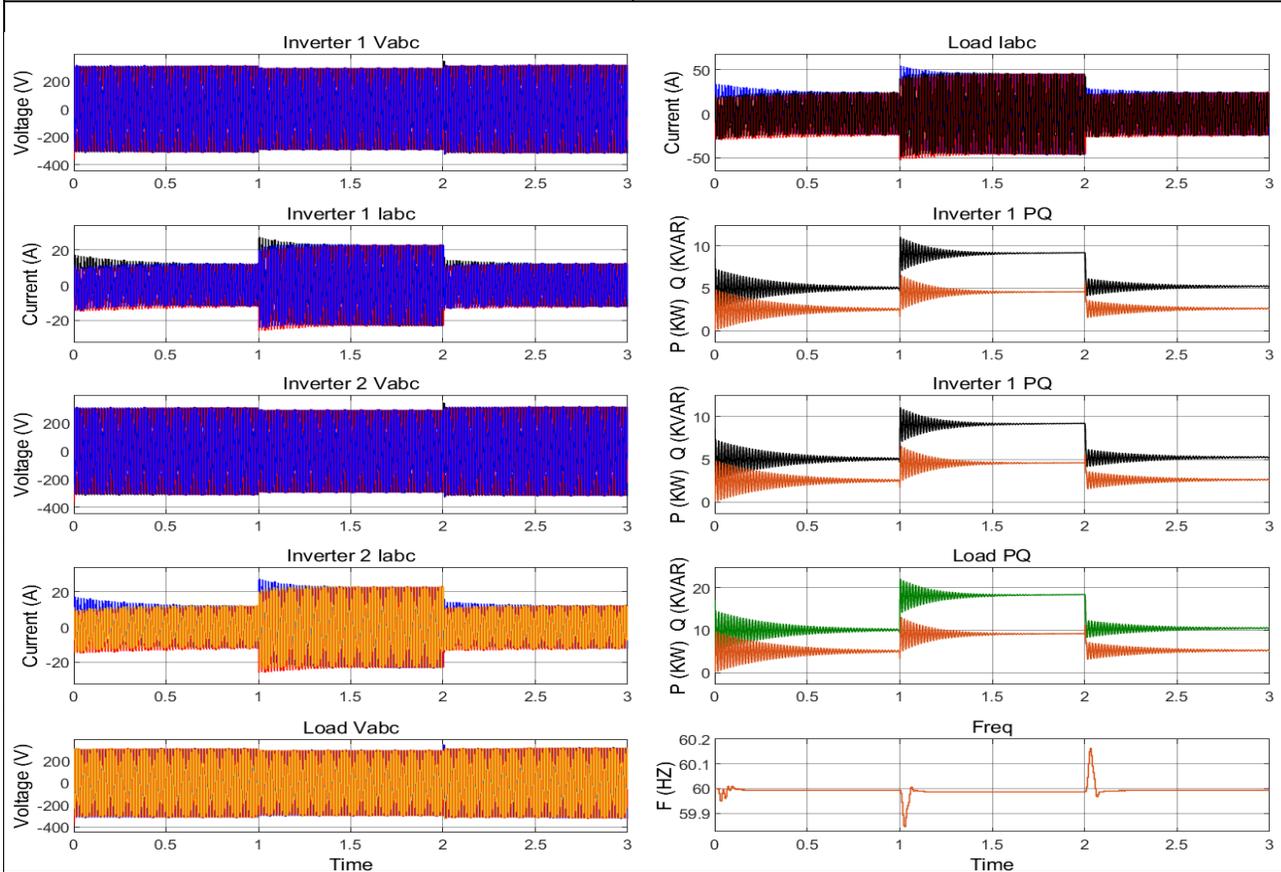

Figure 10: Voltage and current, and power in DG1, DG2 and Load